\documentclass[a4paper, 10pt, conference]{ieeeconf}      

\IEEEoverridecommandlockouts                    
\usepackage{color, soul}
\usepackage{graphicx}

\title{\LARGE \bf Toward Automated Formation of Composite Micro-Structures Using Holographic Optical Tweezers}

\author{Tommy Zhang$^{1}$, Nicole Werner$^{2}$ and Ashis G. Banerjee$^{3}$
\thanks{$^{1}$Tommy Zhang is with the Department of Mechanical Engineering,
        University of Washington, Seattle WA 98195, USA
        {\tt\small tommyz@uw.edu}}%
\thanks{$^{2}$Nicole Werner is with the School of Environmental and Forest Sciences,
        University of Washington, Seattle WA 98195, USA
        {\tt\small nicole84@uw.edu}}%
\thanks{$^{2}$Ashis G. Banerjee is with the Department of Industrial \& Systems Engineering and the Department of Mechanical Engineering,
        University of Washington, Seattle WA 98195, USA
        {\tt\small ashisb@uw.edu}}%
}

\begin{document}

\maketitle
\thispagestyle{empty}
\pagestyle{empty}

\begin{abstract}
Holographic Optical Tweezers (HOT) are powerful tools that can manipulate micro and nano-scale objects with high accuracy and precision. They are most commonly used for biological applications, such as cellular studies, and more recently, micro-structure assemblies. 
Automation has been of significant interest in the HOT field, since human-run experiments are time-consuming and require skilled operator(s). Automated HOTs, however, commonly use 
point traps, which focus high intensity laser light at specific spots in fluid media to attract and move micro-objects. 
In this paper, we develop a novel automated system of tweezing multiple micro-objects 
more efficiently using 
alternate multiplexing formations, such as annular rings and line patterns. Our automated system is realized by augmenting the capabilities of a commercially available HOT with real-time
bead detection and tracking, and wavefront-based path planning.
We demonstrate the usefulness of the system by assembling two different composite micro-structures, comprising 5 $\mu m$ polystyrene beads, using both annular and line shaped traps 
in obstacle-rich environments.
\end{abstract}

\section{Introduction}
Holographic optical tweezers (HOT) enable precise manipulation of an arbitrary number of objects in a micro-environment. The use of a computer controlled spatial light modulator (SLM) to generate holographic patterns allows for the generation of multiple types of optical trap shapes, such as diffraction-limited point traps, annular rings, and line patterns. The ability of HOTs to manipulate multiple micro and nano-scale objects concurrently makes them especially well-suited for biological 
applications. They can function as multi-purpose tools for cellular studies, where they are used extensively for cell transportation, sorting, and characterization 
\cite{ cell_deformation,tug_of_war_tweezers, cell_interactions,cell_translation_rotation,automatic_transportation, singular_perturbation_approach}. Additionally, HOTs have seen a lot of use in the area of micro-assembly, where structures, such as micro-scale fluid pumps, have been assembled \cite{opticalpump}. 

Along with the growth in the use of HOTs, automation has become increasingly important to 
reduce the time and effort to conduct biological studies or assemble structures \cite{banerjee2010,banerjee2012}. Otherwise, trapping and moving the objects
require an experienced operator familiar with the HOT system and the natural limitations of human inputs limits the scalability of concurrent manipulation tasks. 
The augmentation of standard HOT systems with customized hardware and control software has enabled automation of challenging tasks, such as the assembly of packed 3D crystal-lattices using photocured micro-spheres \cite{3dassembly, 3d_assembly2}. In addition, due to the nature of the highly focused laser beam present in HOTs, direct cell manipulation can cause damages, rendering the live samples useless. Methods, which use HOTs as robotic end-effectors by manipulating inert 3D printed or dynamically formed micro-grippers, allow for the safe and efficient transport of living micro-organisms \cite{banerjee2014, indirect_manipulation, microrobots_indirect_manip}. Applications involving in-vivo manipulation of cells are also being investigated using automation methods \cite{invivo}.

The potential applications of HOTs, however, lead to several challenges in automated operations. Since HOTs operate on small-scale objects in fluid media, the random distributions of the objects in the work area often impede the manipulation tasks due to unintentionally trapped objects and/or collisions with the tweezed objects. This stochastic nature of HOT operations requires robust automation system design \cite{Pesce2020OpticalTT}. Additionally, other challenges, such as trap instability and optical misalignment, require methods to compensate for these issues, and ensure that the optical traps are unlikely to lose the manipulated objects.

These challenges are typically addressed through collision-free trajectory planning of the manipulated objects using graph search \cite{rajasekaran2017} or sampling-based techniques \cite{6676857}. However, they only use independently controlled point traps, wherein laser beams are focused at specific spots in the micro-environments. Other multiplexing configurations, such as annular and line traps, have not been used to the best of our knowledge, which could enable more efficient methods for concurrent micro-manipulation. Here, we demonstrate a first-of-its-kind method for automatically assembling composite micro-structures using combinations of optically actuated annular and line traps. In particular, we combine automated object detection with SLM computation and path planning to move user-defined annular and line traps, holding the target objects, to their desired configurations to compose the micro-structures. Experiments with 5 $\mu m$ polystyrene beads show promising outcomes, while also highlighting the need for further improvements in the scalability and robustness of the automation framework.    

\section{Methods}
\subsection{HOT System}
The HOT system is a commercially available kit designed and distributed by Meadowlark Optics \cite{red_tweezers}. The camera used is a Teledyne DASLA HM640 that captures 640 $\times$ 480 pixel monochrome frames at a maximum rate of 300 frames per second. The overall magnification of the camera is 40X based on an oil-immersion 1.35 NA objective. The field of view of the camera is a $120 \ \mu m \times 90 \ \mu m$ rectangular region, which can translate vertically and horizontally by an electronically actuated microscope stage. The SLM is a 512 $\times$ 512 pixel phase-only model from the Boulder Nonlinear Systems with a refresh rate of 100 Hz. The laser used is an IPG Photonics Nd-YAG 5W 1064 nm model with variable power output.

The point traps, annular traps induced by optical vortices, and line traps are generated as follows: 
\begin{enumerate}
    \item \textbf{Point Traps:} Diffraction-limited traps are characterized by high intensity points of light in which small-scale objects are attracted.
    
    \item \textbf{Annular Traps:} Optical traps with a non-zero topological charge, carry orbital angular momentum and are classified by a doughnut shaped high intensity distribution around a dark, zero intensity center point in the focal plane. These annular traps are capable of manipulating metallic, reflecting, and absorbing objects, which could not be trapped using conventional point traps \cite{grier2003revolution}. In addition, traps with a topological charge are capable of trapping dielectric particles in the high intensity distribution outer ring, giving rise to ring shaped constructs of micro-beads. They have been the subject of research in applications, such as micro-fluidic pumps and optical sorting of particles \cite{opticalpump,opticalsorting}. 
    
    \item \textbf{Line Traps:} In our HOT system, line traps are modified from \cite{roichman2006projecting}, in which shape-phase holography is used to generate arbitrary potential energy wells along 1D curves. In our application, these curves are linear and can be of any arbitrary length and direction. 
\end{enumerate}

\begin{figure}[!ht]
    \includegraphics[scale=0.125]{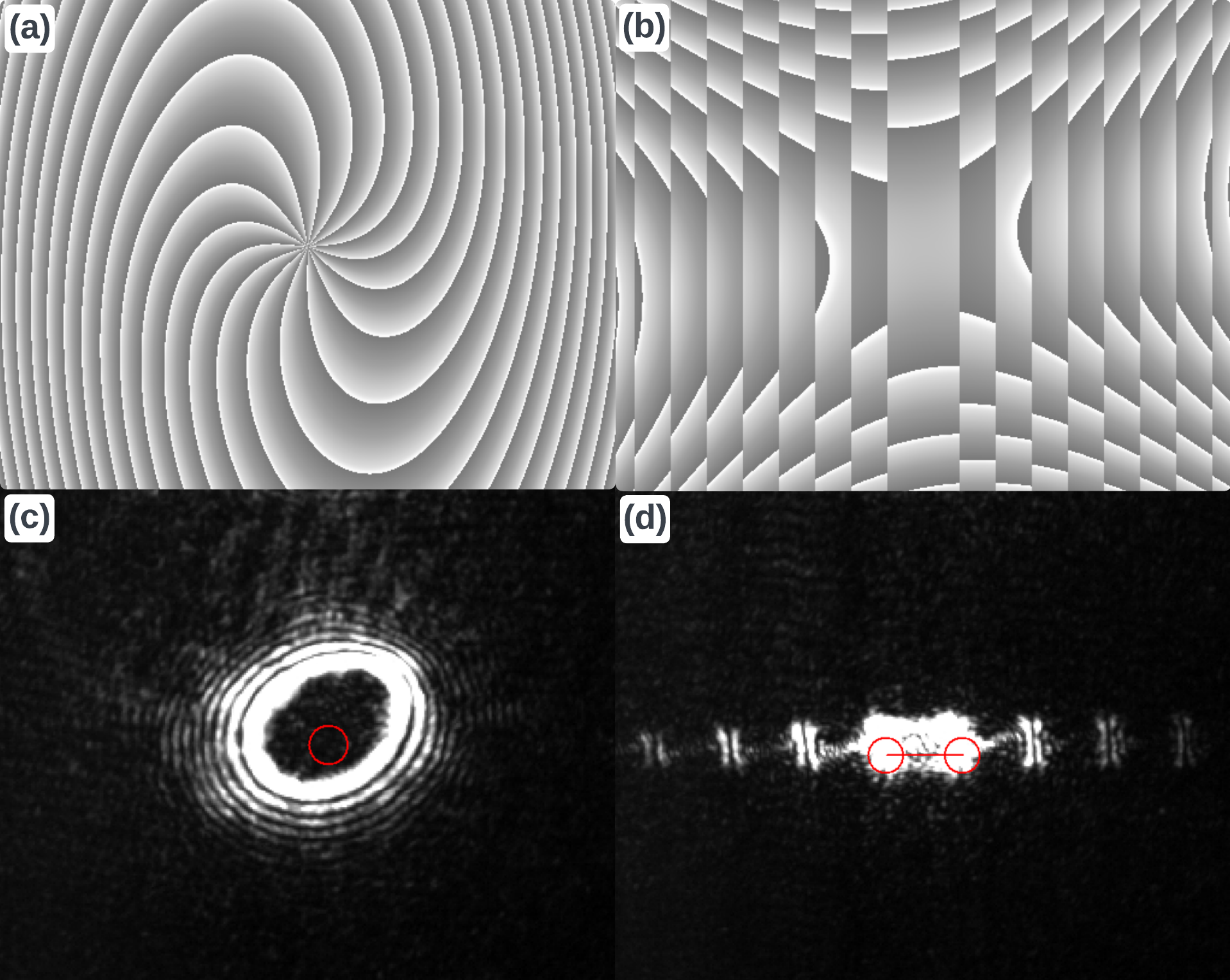}
    \caption{Phase mask and light intensity distribution on the focal plane. Annular trap phase mask \textbf{(a)}, line trap phase mask \textbf{(b)}, annular trap light intensity distribution using a topological charge $l=15$ \textbf{(c)}, line trap light intensity distribution \textbf{(d)}.}
    \label{fig:phasemask}
\end{figure}

Fig. \ref{fig:phasemask} shows the corresponding phase and light intensity distributions for the annular and line traps.

\subsection{Problem Formulation}
\begin{figure}[!ht]
    \includegraphics[scale=0.2]{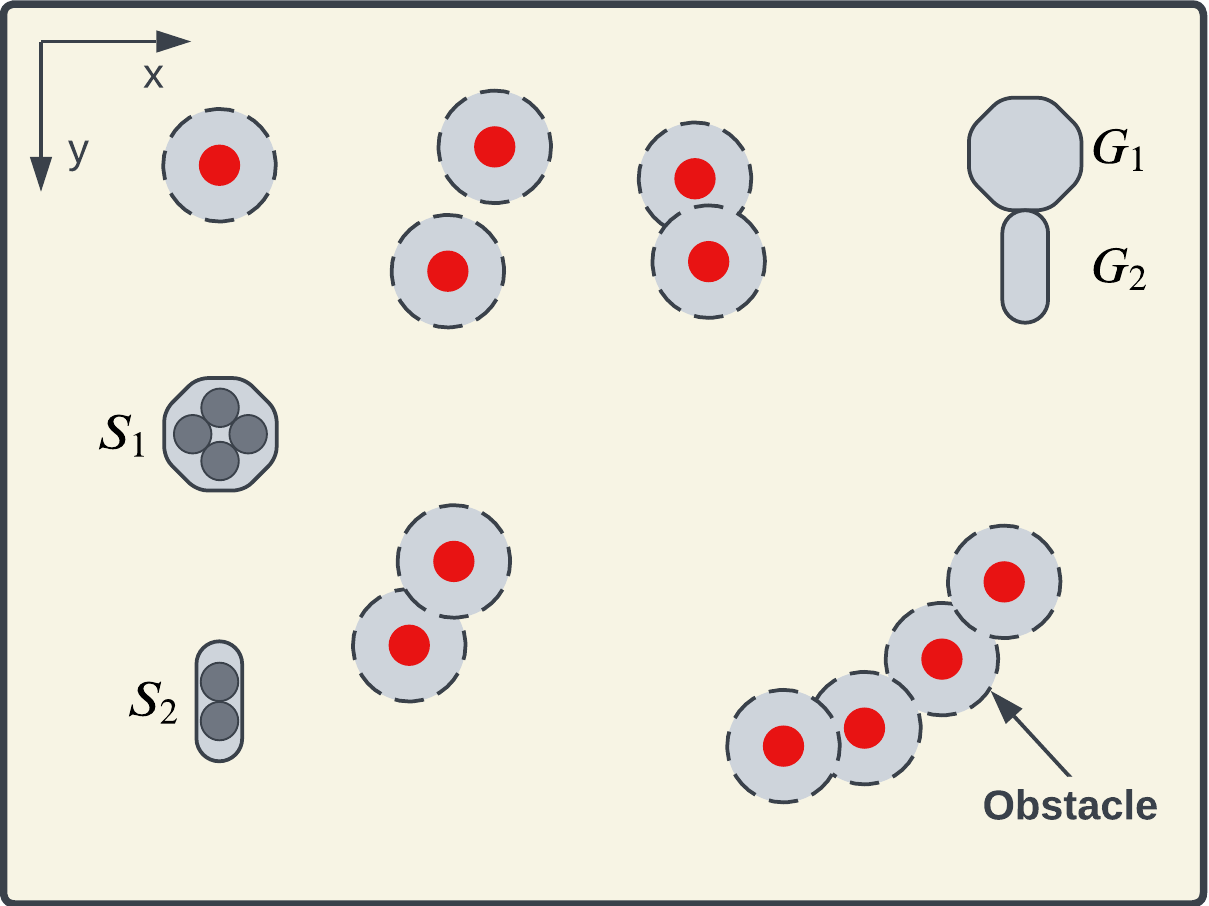}
    \caption{Illustration of the problem formulation. $S_i$ and $G_i$ denote the start and goal configurations, respectively. Red circles denote the obstacles with buffer regions around them. Gray beads indicate the objects of interest for path planning.}
    \label{fig:prob_form}
\end{figure}

In our study, we aim to move a set of multiplexed optical traps from their starting states to user-specified goal states while avoiding the obstacles in the micro-environment. We denote the initial and goal sets as $S^{a} = \{S^{a}_{1}, S^{a}_{2}, ..., S^{a}_{n}\}$, $S^{l} = \{S^{l}_{1}, S^{l}_{2}, ..., S^{l}_{m}\}$, $G^{a}= \{G^{a}_{1}, G^{a}_{2}, ..., G^{a}_{n}\}$, $G^{l}= \{G^{l}_{1}, G^{l}_{2}, ..., G^{l}_{m}\}$, where $S^{a}$ and $G^{a}$ contains $n$ annular traps and $S^{l}$ and $G^{l}$ contains $m$ line traps. The set of $k$ obstacles, denoted by $O = \{O_1, O_2,..., O_k \}$ are unknown before runtime and assumed to only consist of other micro-beads. 

This can be formulated as a path planning and control problem, where the optical trap trajectories are planned and executed simultaneously. A valid solution would be one which, starting with $m$ line traps and $n$ annular traps, would generate trajectories from all the starting states $S^{a}$ and $S^{l}$, to all the corresponding goal states $G^{a}$ and $G^{l}$. Our experiments demonstrate this problem with two multiplexed traps, located at $S_1$ and $S_2$, and generate trajectories to the goal locations $G_1$ and $G_2$, while avoiding the obstacles in $O$, as shown in Fig. \ref{fig:prob_form}.

\subsection{Technical Approach}
By developing a HOT automation framework in C++ to realize automatic control of micro-particles, we are able to achieve a fast and reliable method for constructing scalable composite shapes using inert polystyrene beads. The framework consists of various modules as follows:

\subsubsection{Bead Detection}
A blob detection based image processing algorithm is used to detect and output the coordinates of all the micro-beads in the camera field of view. The algorithm is a modified version of that described in \cite{ekta}, where only the histogram equalization and blob detection methods are used to identify the polystyrene beads using the OpenCV package. This method enables us to quickly and accurately capture the locations of each individual detected bead in order to guide the other algorithms in the automation system.

\subsubsection{SLM Communication}
The phase patterns for the SLM are generated using the Meadowlark Optics hologram engine software, which accepts user datagram protocol (UDP) packets as inputs to calculate and display the intended phase mask on the SLM. The phase mask calculations are done locally on the GPU, which allows for fast refresh rates when changing the trap parameters or the number of traps.

\begin{figure}[th!]
    \includegraphics[scale=0.2]{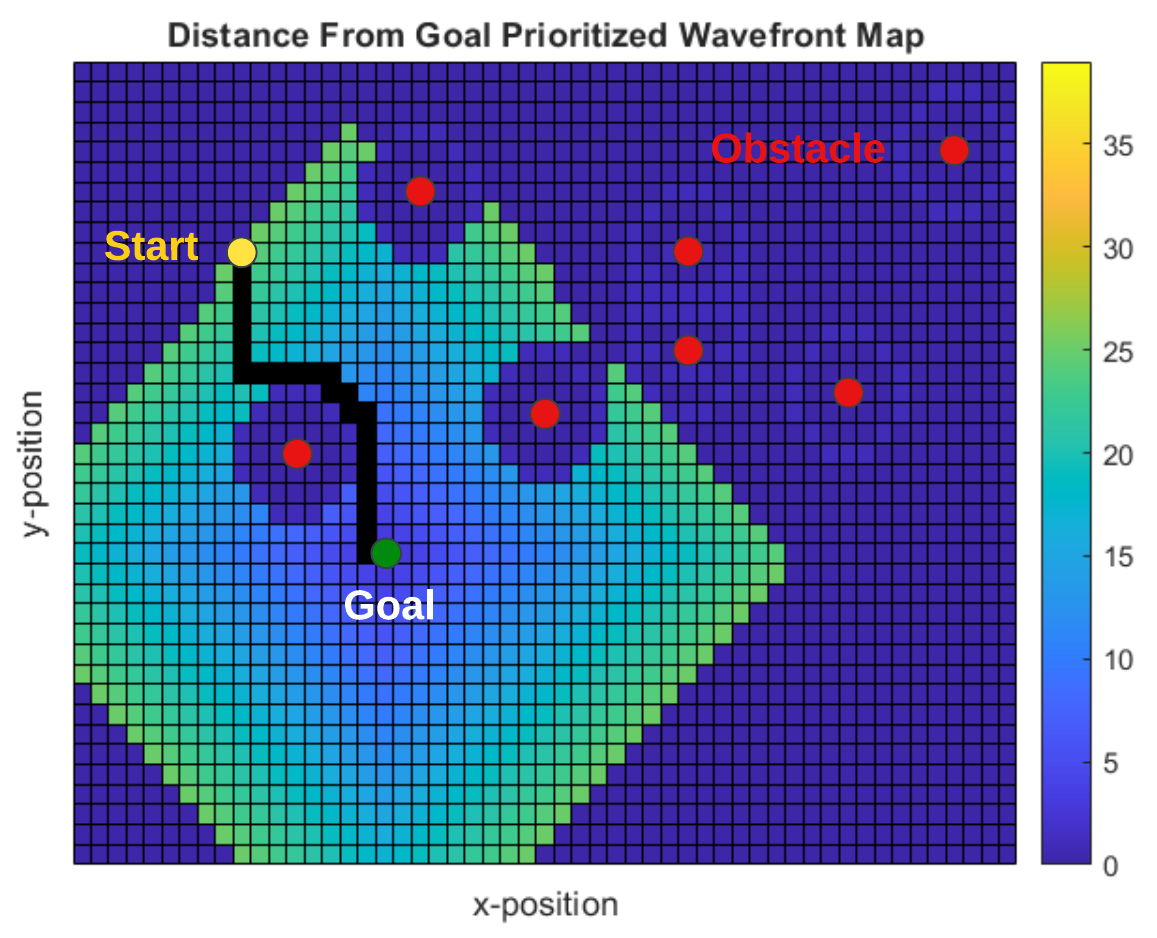}
    \caption{Heatmap of the primary wavefront planner to construct a P-shaped micro-structure. Yellow and green dots represent the start and goal locations, respectively. Red dots represent the obstacle beads and solid blue regions denote the obstacle buffer zones and unexplored areas.}
    \label{fig:wf_planner}
\end{figure}

\begin{figure}[th!]
    \includegraphics[scale=0.2]{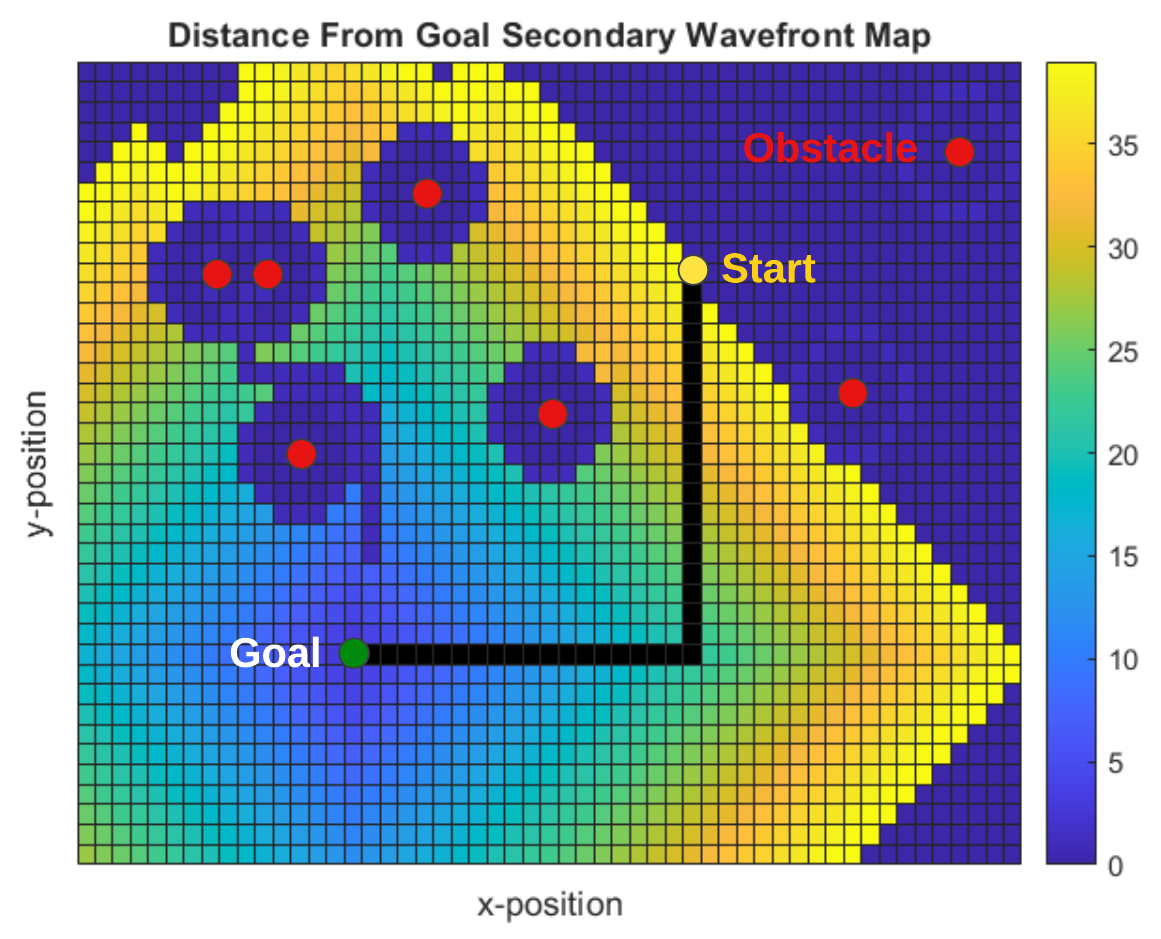}
    \caption{Heatmap of the secondary wavefront planner to construct a P-shaped micro-structure. Note the addition of the first path as obstacles.}
    \label{fig:sec_wf_planner}
    \end{figure}

\subsubsection{Path Planning}
In order to generate collision-free trajectories to assemble micro-structures, we use an artificial potential field-based path planning algorithm. Specifically, the wavefront expansion planning algorithm which finds a path from a specified starting state to a goal state. The wavefront planner is advantageous in that it always finds a suitable path if one exists, and does not get trapped in local minima, unlike other implementations of potential field-based path planners \cite{LaValle_2006}. Our wavefront planner works in a discretized gridspace with Von Neumann neighborhood connectivity. It uses breadth-first search from the goal state to recursively traverse the nodes within a target Manhattan distance until the start state is reached. This target distance is based on the size of the gridspace, and is defined as the length of an edge between any two neighboring nodes.

The primary (prioritized) and secondary wavefront expansion results for a composite P-shaped structure formation experiment are shown in Fig. \ref{fig:wf_planner} and Fig. \ref{fig:sec_wf_planner}, respectively. The gridspace is initialized to all zeros and the goal state is initialized to a value of two. The obstacles are identified using the bead detection algorithm and added to the gridspace with a buffer region of radius $r_{br} = 3\mu m$. Once the breadth-first search algorithm finds the start state, the algorithm backtracks from the start state based on the maximum decrease in distance from the goal. An implementation of this algorithm with gridspace dimensions matching the camera image size is inefficient for run-time computation. Therefore, we downsample to a 53 $\times$ 40 gridspace, which in our case corresponds to 2.25 $\mu m$ per pixel. This enables a balance of computation speed and granularity, with each pixel (approximately) corresponding to the radius of the used microbeads.

We then perform a linear interpolation of the downsampled points in the wavefront gridspace to generate a smooth path while executing the trap motions. This is achieved by generating intermediate points at  
every $0.5 \ \mu m$ interval along the planned path. Since the trap speed is controlled using delays between discrete movements, a sufficiently high number of interpolated points is required for low-speed motion. The delay between the trap movements is calculated as $\frac{1}{2} v$, where $v$ is the desired trap speed in $\frac{\mu m}{s}$.

\subsection{Execution}
The experiments are executed by integrating the various automated subsystems together to assemble the composite shapes. This execution is achieved as follows:

\begin{enumerate}
    \item \textbf{Trap Generation:} The annular and line traps are first generated by the user, and set at their desired starting positions $S_1$ and $S_2$. 
    \item \textbf{Obstacle Detection:} All the beads in the camera's field of view are first detected and stored. The obstacles are determined by taking the set of trapped beads and removing them from the set of detected beads. These beads are not intended to be trapped and are all considered as obstacles. They are then added to the set $O$.
    \item \textbf{Priority Path Planning:} The wavefront planner is executed on the prioritized path first. The planner outputs a valid path $P_1$ from $S_1$ to $G_1$ while avoiding the obstacles in $O$. The path is then smoothed by interpolating in between the coarse planning grid points. 
    \item \textbf{Secondary Path Planning:} The obstacles of the second path are appended with the prioritized path. This ensures that the trajectories of both the beads do not intersect and cause potential collisions between the beads. The wavefront planner is again executed and the path is smoothed. The output is a smoothed path $P_2$ that avoids the obstacles defined by $O \cup P_1$ from $S_2$ to $G_2$.
    \item \textbf{Obstacle Trapping:} All the obstacles $O$ in the original set are then trapped. This reduces the likelihood of possible obstacle collisions and allows the obstacles to be treated as stationary objects. Note this step is not executed for the experiment in Fig. \ref{fig:flower_expt}.
    \item \textbf{Path Execution:} Both the planned paths are executed simultaneously. The beads are translated point to point following the smoothed trajectory. A user-specified delay is added between each move to control the trap velocity.
\end{enumerate}

\section{Experimental Results}
\subsection{Experimental Setup}
The micro-bead samples are prepared using 5$\mu$m polystyrene beads manufactured by Bangs Laboratories. The stock solution is diluted with DI water to achieve the desired bead density for the experiments. We observe that the exact dilution ratio is not critical for achieving a reasonable bead density. Instead, the time elapsed during experimentation with an activated laser contributes to the desiccation of the bead solution. $10 \mu L$ solutions of diluted polystyrene beads are used for each individual experiment and deposited onto a glass bottomed petri dish. 

The annular traps require tuning the topological charge $l$ and z-offset based on the number of trapped beads. If these values are too small or large, 
the annular traps become unstable, leading to the beads becoming untrapped or optically bound in an out-of-focus plane \cite{opticalbinding}. Additionally, the line traps are experimentally tuned to specific lengths based on the number of beads trapped. If the line trap is too short for the number of trapped beads, interference and stacking of beads occur causing trap instability. 
On the other hand, if the line trap is too long for the number of trapped beads, semi-stable regions within the line trap leads to uneven spacing of the beads. Additionally, unnecessarily long line traps attract and trap other nearby beads unintentionally.

\subsection{Results}
Our experiments aim to demonstrate the capability of our system to automatically assemble a composite micro-structure of various shapes with random obstacle positions in a timely manner. The experiments consist of an annular trap, which is formed by the user in a predetermined position consisting of four micro-beads clumped closely together. Similarly, the line traps consist of either two or three beads, which are trapped by the user in a predetermined location. The start and goal locations of the traps are based on the desired composite shape. Ideally, the annular and line traps should yield trajectories that avoid all the obstacles. However, the stochastic nature of the environment results in the beads diffusing in from the sides of the workspace and into the focal plane, limiting the repeatability of the experimental scenarios.

\begin{figure}[h]
    \includegraphics[scale=0.15]{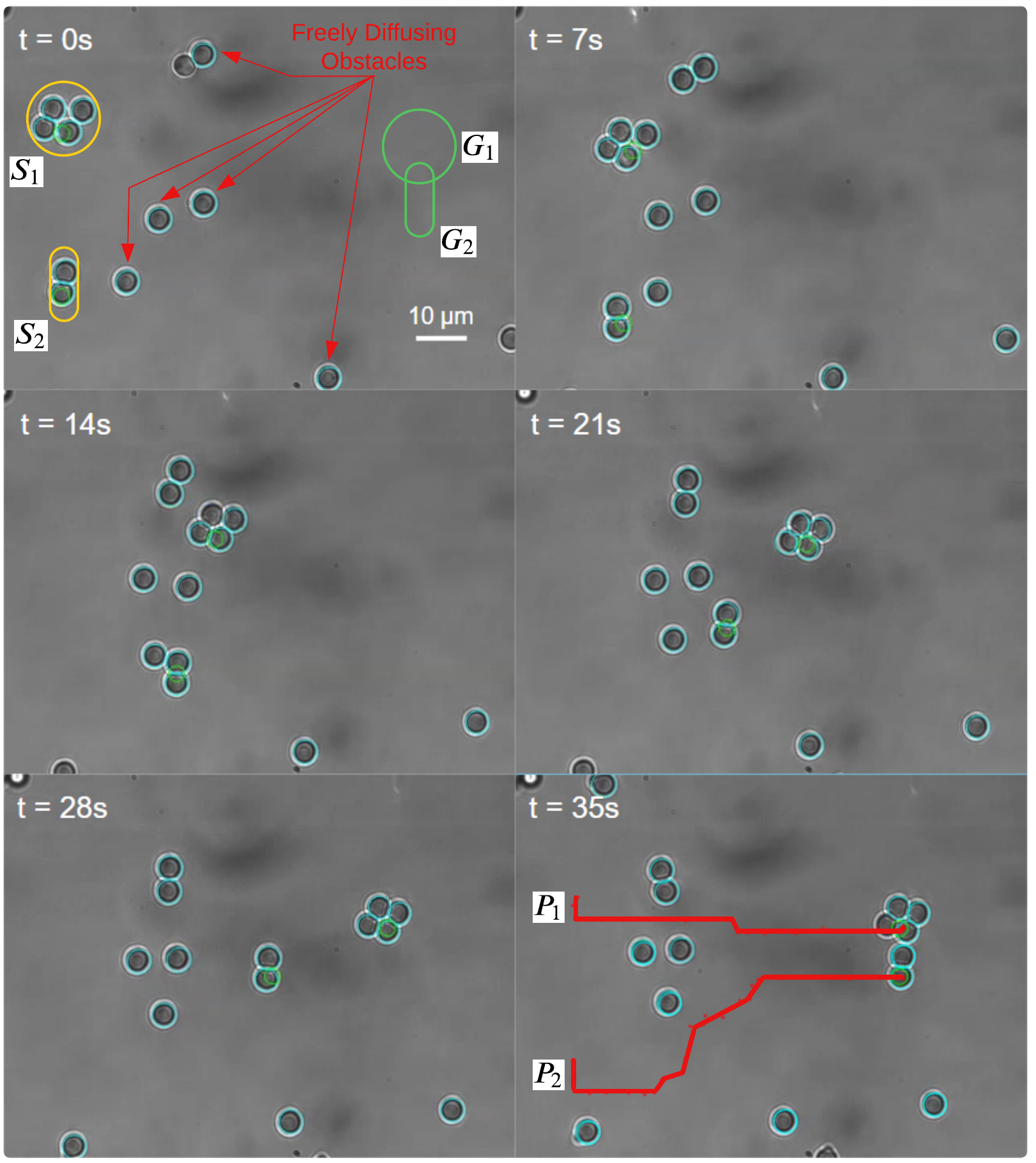}
    \caption{Time lapse images of a flower-shaped structure generation using two multiplexed optical traps. The time elapsed is 35 seconds and the frames are taken in 7 second intervals. The green circles are active traps and the blue circles are the detected beads.}
    \label{fig:flower_expt}
\end{figure}

Our first experiment is on assembling a flower-shaped structure using an annular trap consisting of four beads and a line trap comprising two beads. The starting points are placed near the left side of the workspace $45 \mu m$ apart. The goal positions are near the right side of the workspace, and the annular and line traps are kept close together. Time-lapse images of this experiment are shown in Fig. \ref{fig:flower_expt}.
The obstacle beads are allowed to undergo free diffusion. The structure is assembled in 35 seconds with both the trap speeds set to 1.5 $\frac{\mu m}{s}$. We observe that the line trap and obstacle bead do come in close proximity at the in 14 second mark; however, they are separated at the 21 second mark and the experiment continues as planned. Although the motion planner provides a buffer region around the obstacles, spurious light intensity distributions around the obstacle sometimes moves it further than expected due to Brownian motion. These observations guide our next experiment, where all the obstacles are trapped to prevent them from moving.

\begin{figure}[h]
    \includegraphics[scale=0.15]{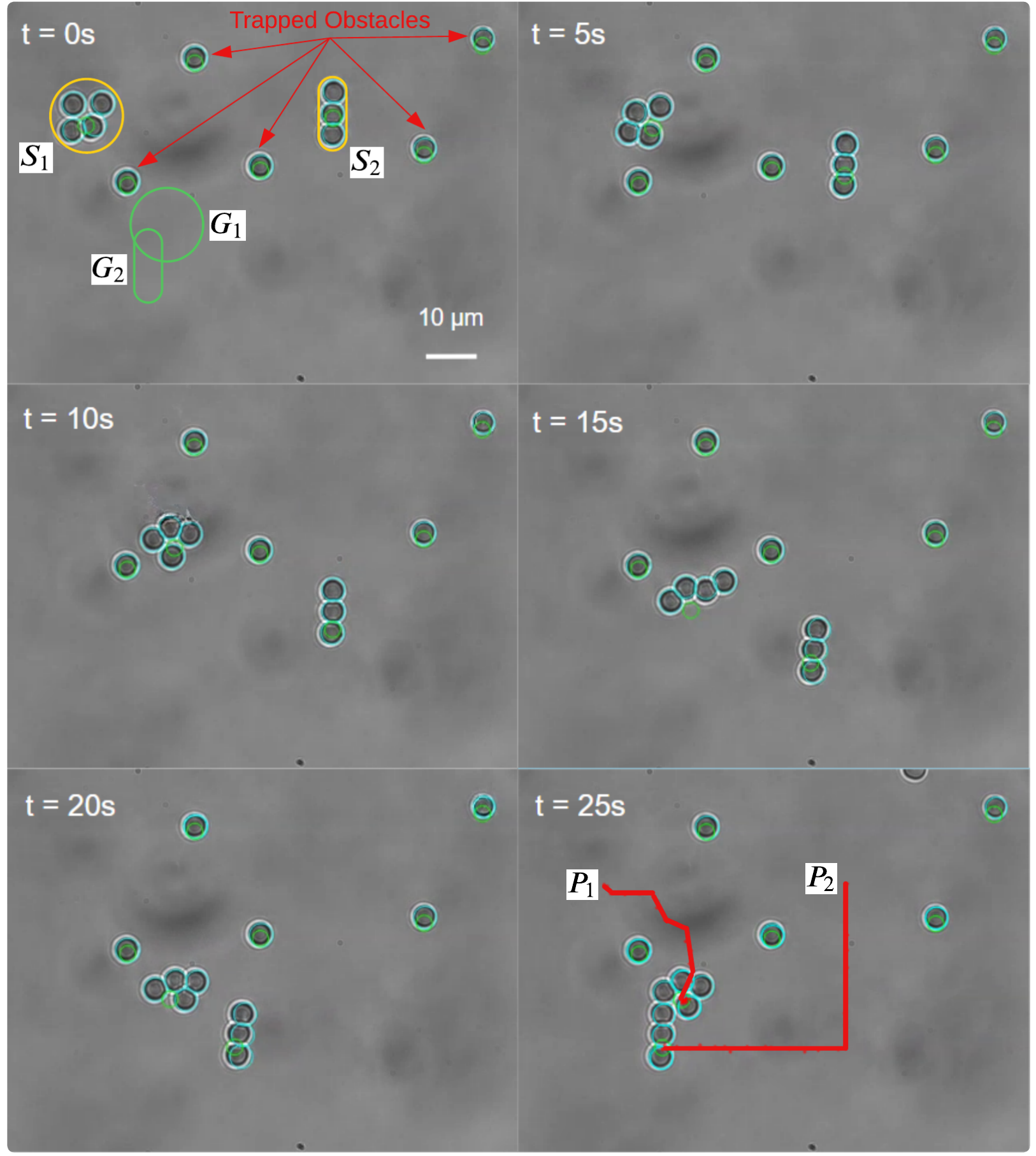}
    \caption{Time lapse images of a P-shaped structure generation using two multiplexed optical traps. The time elapsed is 25 seconds and the frames are taken in 5 second intervals. The green circles are active traps and the blue circles are the detected beads.}
    \label{fig:P_expt}
\end{figure}

The second experiment, shown in Fig. \ref{fig:P_expt}, demonstrates the formation of a letter P-shaped structure. The structure consists of a line trap with three beads and an annular trap with four beads. Both the traps start near the upper region of the workspace and the goal is to assemble the structure in the lower left region. The structure is assembled in 25 seconds, with trap speeds being the same at 1.5 $\frac{\mu m}{s}$. In this case, we observe that the traps can avoid the obstacles with a sufficient clearance to prevent unintended trapping by circumventing the attractive forces from the neighboring traps.

Additionally, we record the run times of the various execution modules. 
As shown in Table \ref{tab:my_table}, the loop times of all the modules are completely suitable for a real-time automated system. Even the most (computationally) limiting module of bead detection runs at an average rate of 113 Hz. These rates are very promising in terms of enabling future expansion of the system to a greater number of multiplexing formations, larger workspaces with a denser and heterogeneous collection of objects, and implementation of simultaneous trajectory updates during run time. 

\begin{table}[htbp]
  \caption{Run Time Values for the Different Automation Components}
  \centering
  \begin{tabular}{|c|c|c|}
    \hline
    \textbf{Module} & \textbf{Loop Time (in ms)} & \textbf{Rate (Hz)} \\
    \hline
    Bead Tracking & 0.49 $\pm$ 0.210 & 2040 \\
    \hline
    Bead Detection & 8.86 $\pm$ 0.611 & 113 \\
    \hline
    Path Planning & 7.68 $\pm$ 1.37 & 130 \\
    \hline
    SLM Communication & 0.086 $\pm$ 0.029 & 11664 \\
    \hline
  \end{tabular}
  \label{tab:my_table}
\end{table}

\section{Discussion}

Issues arise when using polystyrene micro-beads in a fluid medium, such as over saturation that causes an uncontrollable and unavoidable clustering due to thermal fluctuations, Brownian motion, and other variables \cite{temp_change_laser}. Even with a low bead density, micro-beads from outside the field of view can diffuse into the workspace to interfere with the trapped beads or present themselves as obstacles during planned trajectories. This stochastic nature of experiments involving HOTs is challenging and requires active control measures to mitigate the risk of failure.   

Our first experiment demonstrates the formation of a flower-like structure without any active trapping of the obstacles. We notice that one of the obstacles loosely follows the line trap trajectory for a short period of time, bringing it outside the buffer region considered by the path planner. The assumption that the micro-beads would only exhibit Brownian motion during the experiment is, therefore, violated. We believe this behavior stems from imperfections and misalignment in the HOT system. 

The second experiment employs a simple, but effective strategy of actively trapping all the detected obstacles. While this strategy is advantageous for our experiments, it does come with some possible drawbacks. Since laser power is divided among all the optical traps based on the trap type, trapping all the obstacles effectively reduces the overall power provided to the annular and line traps. This is, in fact, observed in the experiment, when the annular trap almost loses a bead during transport, as shown in Fig. \ref{fig:P_expt}, at the 15 second time stamp. In future experiments, dynamically allocating the power based on the number of obstacles would alleviate this issue but bring us closer to the maximum operating power of the laser. 

The wavefront potential field planner tends to yield paths that are close to the obstacles. 
Other variations of the potential field method do not exhibit this undesirable behavior, and could be explored to improve the robustness of the path planning algorithm. Assuming the obstacles are trapped, an algorithm that provides a better approach of handling obstacles could reduce the buffer region radius and generate feasible trajectories even for a high density of obstacles in the workspace. 

The prioritized wavefront planner is not limited to two simultaneous trajectories as shown in the experiments, and can be expanded to yield a larger number of trajectories. However, such expansion is expected to decrease the likelihood of finding feasible solutions due to the consideration of other trajectories as obstacles. Since criss-cross trajectories are not possible with this planner, improved performance with a large number of traps could be achieved in the future using adaptive motion planning and control methods.

\section{Conclusions}
In this paper, we provide the first step toward automated assembly of various composite micro-structures using multiplexed optical traps. Specifically, annular and line traps are used to trap multiple beads, and a potential field-based path planner is developed to plan the trajectories of the trapped beads while avoiding randomly distributed obstacles. This enables efficient transport of micro-particles while minimizing the computational cost of path planning. It also highlights the ability of our HOT system to transfer a predefined set of particles to various user-defined goal destinations. 

In the future, we plan to consider various ways to 
expand upon and generalize the strategies we developed in this work. Our current experiments only use 5$\mu m$ beads for micro-structure assembly. However, other bead sizes, or a heterogeneous mixture of beads, might be needed for certain assemblies. 
Since smaller beads exhibit greater Brownian motion, we may not be able to track or control the particles using our existing algorithms. A more involved model predictive controller \cite{banerjee2018} might be required to handle the fast, stochastic motion of the smaller particles. 
Additionally, our HOT system is capable of controlling the z-coordinates of the traps, independent of their XY locations. This would allow us to construct 3-dimensional structures using different multiplexed formations. Future work on this topic would require hardware modifications and algorithm refinements to allow for imaging and tweezing of the micro-beads along different planes.

\bibliographystyle{IEEEtran}
\bibliography{citations}


\end{document}